\documentclass[aps,prd,nofootinbib,amsmath,amssymb,superscriptaddress,twocolumn,10pt]{revtex4}

\usepackage{txfonts}
\usepackage{amsmath}
\usepackage{graphicx}
\usepackage{dcolumn}
\usepackage{bm}
\usepackage{amssymb}
\usepackage{latexsym}

\newcommand{\be}{\begin{equation}}
\newcommand{\ee}{\end{equation}}
\newcommand{\bq}{\begin{eqnarray}}
\newcommand{\eq}{\end{eqnarray}}
\begin{document}

\title{Confronting brane inflation with \textit{Planck} and pre-\textit{Planck} data}

\author{Yin-Zhe Ma}
\email{mayinzhe@phas.ubc.ca} \affiliation{Department of Physics
and Astronomy, University of British Columbia, Vancouver, V6T 1Z1,
BC, Canada} \affiliation{Canadian Institute for Theoretical
Astrophysics, Toront, M5S 3H8, Ontario, Canada.}
\author{Qing-Guo Huang}
\email{huangqg@itp.ac.cn} \affiliation{State Key Laboratory of
Theoretical Physics, Institute of Theoretical Physics, Chinese
Academy of Sciences, Beijing 100190, China}
\author{Xin Zhang}
\email{zhangxin@mail.neu.edu.cn} \affiliation{College of Sciences,
Northeastern University, Shenyang 110004, China}\affiliation{Center
for High Energy Physics, Peking University, Beijing 100080, China}

\begin{abstract}
In this paper, we compare brane inflation models with the
\textit{Planck} data and the pre-\textit{Planck} data (which
combines \textit{WMAP}, ACT, SPT, BAO and $H_{0}$ data). The
\textit{Planck} data prefer a spectral index less than unity at more
than $5\sigma$ confidence level, and a running of the spectral index
at around $2\sigma$ confidence level. We find that the KKLMMT model
can survive at the level of $2\sigma$ only if the parameter $\beta$
(the conformal coupling between the Hubble parameter and the
inflaton) is less than $\mathcal{O}(10^{-3})$, which indicates a
certain level of fine-tuning. The IR DBI model can provide a
slightly larger negative running of spectral index and red tilt, but
in order to be consistent with the non-Gaussianity constraints from
\textit{Planck}, its parameter also needs fine-tuning at some level.
\end{abstract}

\maketitle

\section{Introduction}
The ongoing astronomical observations, such as \textit{WMAP}
\cite{Hinshaw12}, \textit{Planck}
\cite{Planck13-1,Planck13-16,Planck13-22}, SDSS \cite{Anderson13},
ACT \cite{Sievers13} and SPT \cite{Hou12}, have been measuring the
cosmic microwave background (CMB) and large scale structure to an
unprecedented precision. This provides an excellent opportunity to
probe the physics in the early Universe with the underlying
fundamental theories. One of the leading candidates of generating
initial fluctuations in the early Universe is inflation
\cite{Guth81,Linde82}. The inflation paradigm offers a compelling
explanation for many puzzles in the standard hot big-bang
cosmology, such as the flatness problem, homogeneity problem and
horizon problem \cite{Guth81}. The accelerated expansion period in
the early Universe provides a nearly scale-invariant primordial
power spectrum which has already been supported by the
measurements of CMB anisotropy
\cite{Hinshaw12,Sievers13,Hou12,Planck13-1,Planck13-16,Planck13-22,Planck13-24}.
In spite of its phenomenological success, inflation remains a
paradigm rather than a fundamental theory, which in principle can
be implemented by various models from different microscopic
physical constructions \cite{Bean08}. The fact that it is easy to
construct a wide variety of inflation models does not mean that
any of them will turn out to be the true mechanism.
Actually, effective field theory models of inflation should by
definition be understood as valid only up to some energy scale that
is low enough, and so the singularity problem and any
``trans-Planckian'' effects are out of the range of validity of the
models~\cite{Starobinsky80,Martin01}. If one would like a UV
completion to any effective field theory ideas, one might hope that
the string theory would provide such a way. Undoubtedly, inflation
can be successfully realized in a string context.

The string inflation model considered in this paper is the brane
inflation scenario, proposed in \cite{Dvali99,Tye08} originally,
which offers a class of observational signatures. In this
scenario, the inflation is driven by the potential between the
parallel dynamic brane and antibrane
\cite{Dvali01,Burgess01,Quevedo02}, and the distance between the
branes in the extra compactified dimensions plays the role of the
inflaton field. This inflation scenario can be realized via two
viable mechanisms, namely, the slow-roll and Dirac-Born-Infeld
(DBI) inflations \cite{Bean08}.

The original brane inflation model is the slow-roll inflation
model \cite{Dvali99,Burgess01,Quevedo02,Dvali01} where branes and
antibranes are slowly moving towards each other in a flat
potential. 
The KKLMMT model~\cite{Kachru03} provides such an example. In this
model, the antibrane is fixed at the bottom of a warped throat,
while the brane is mobile and experiences a small attractive force
towards the antibrane \cite{Kachru03,Ma09}. When the brane and
antibrane collide and annihilate, the inflation ends and the hot
big-bang epoch is initiated. The annihilation of the brane and
antibrane makes the universe settle down to the string vacuum
state that describes our Universe \cite{Kachru03,Ma09}. For
extensive studies on the KKLMMT model and other types of slow-roll
brane inflation models, see Refs.
\cite{Firouzjahi05,Baumann06,Huang06,Huang:2006zu,Zhang06,Bean08,Ma09,Baumann08,Baumann10}.

Another inflationary mechanism is the DBI inflation. In this
paradigm, the speed of the rolling brane is not determined by the
shape of the potential but by the speed limit of the warped
spacetime
\cite{Silverstein:2003hf,Alishahiha:2004eh,Chen:2004gc,Chen:2005ad,Chen:2005fe}.
The warped internal spaces naturally arise in the extra dimensions
due to the stabilized string compactification.

In order to test the inflationary paradigm and explore the
dynamics of the internal space, we will scan the parameter spaces
of these two types of inflation models subject only to the
requirement that they provide enough \textit{e}-folding number to
solve the flatness, horizon and homogeneity problems. This is
because solving the problems of standard cosmology is the basic
motivation of the inflation paradigm and the most attractive
feature of inflation models \cite{Bean08}. Then we will explore
the observational signatures that are allowed by brane inflation
dynamics and constrain the model parameters with the current
observational CMB data. We will see that the current observational
data are able to tighten up the parameter space of brane inflation
to a great extent and the generic models need to be fine-tuned to
match the current observations.

Recently the \textit{Planck} team just released the results from
the $2.7$ full-sky surveys \cite{Planck13-1}. For the $\Lambda
$CDM model, \textit{Planck} data combined with \textit{WMAP}
polarization data (hereafter \textit{Planck}+ WP) show that the
index of the power spectrum satisfies
\cite{Planck13-16,Planck13-22}
\begin{equation}
n_{s}=0.9603 \pm 0.0073 ~~~(1\sigma~ \text{CL}), \label{nsnumber1}
\end{equation}
at the pivot scale $k_0=0.05$ {Mpc}$^{-1}$, which rules out the
exact scale invariance ($n_{s}=1$) at more than $5\sigma$. If the
running of  spectral index $\alpha_s=d n_{s}/d \ln k$ is released
as a free parameter, the spectral index becomes redder,
\begin{equation}
n_{s}=0.9561 \pm 0.0080 ~~~(1\sigma~ \text{CL}), \label{nsnumber2}
\end{equation}
while the running of the spectral index is not equal to zero at less
than $2\sigma$ CL,
\begin{equation}
d n_{s}/d \ln k= -0.0134 \pm 0.0090 ~~~(1\sigma~ \text{CL}).
\label{asnumber1}
\end{equation}

For a comparison, in \cite{Cheng13}, we combined the \textit{WMAP}
9-year data \cite{Hinshaw12} with ACT data \cite{Sievers13}, SPT
data \cite{Hou12}, as well as BAO data
\cite{Anderson13,Beutler11,Blake12} and $H_{0}$ data
\cite{Riess11} (hereafter, we call this combined data set the
``\textit{WMAP}9+'' data set), and we obtained a red spectral
index of power spectrum at the pivot scale $k_{0}=0.002$
{Mpc}$^{-1}$,
\begin{equation}
n_{s}=0.961 \pm 0.007~~~ (1\sigma ~ \text{CL}). \label{nsnumber3}
\end{equation}
But if we let the running of the spectral index be $\alpha_s=d
n_{s}/d \ln k$ as a free parameter, the spectral index becomes
\begin{equation}
n_{s}=1.018 \pm 0.027~~~(1\sigma ~ \text{CL}), \label{nsnumber4}
\end{equation}
and the running of the spectral index becomes
\begin{equation}
\alpha _{s}=dn_{s}/d\ln k=-0.021 \pm 0.009 ~~~(1\sigma~\text{CL}).
\label{runnumber1}
\end{equation}

In addition, the joint constraints on $r$ (tensor-to-scalar ratio)
and $n_{s}$ already become a very sensitive tool to constrain
inflation models. In \cite{Hinshaw12}, it is found that inflation
models with power-law potential $\phi^{4}$ cannot provide a
reasonable number of \textit{e}-folds (between $50$--$60$) in the
restricted space of $r$--$n_{s}$ at around $2\sigma$ level.
Reference \cite{Cheng13} pushes this limit further and shows that
with the combination of \textit{WMAP}9, ACT, SPT, BAO and $H_{0}$
data, the inflation potential with power law form $\phi^{p}$ can
only survive if $p$ is in the range of $0.9$--$2.1$.

Besides the above conventional parameters that have been used to
constrain inflation models, \textit{Planck} data are also able to
constrain the non-Gaussianity of primordial fluctuations. The
\textit{Planck} found that the local, equilateral and orthogonal
types of non-Gaussianity are
\begin{eqnarray}
f^{\rm{local}}_{\rm{NL}}= 2.7 \pm 5.8 ~~~(1\sigma ~ \text{CL}),
\nonumber \\
 f^{\rm{equil}}_{\rm{NL}}= -42 \pm 75 ~~~(1\sigma ~
\text{CL}), \nonumber \\
f^{\rm{orth}}_{\rm{NL}}= -25 \pm 39 ~~~(1\sigma ~ \text{CL}).
\label{nGnumber}
\end{eqnarray}
These place very tight constraints on the inflation model space.

Based on the \textit{WMAP} 3-year and 5-year results, Refs.
\cite{Huang06} and \cite{Ma09} investigated brane inflation models
and showed that the KKLMMT model cannot fit \textit{WMAP}+SDSS
data at the level of one standard deviation and a fine-tuning (at
least one part in a hundred) is needed at the level of two
standard deviations. Since the CMB data have been dramatically
improved over the past several years, it is meaningful to see how
the status of brane inflation is affected by the arrival of the
new CMB data, especially the \textit{Planck} and \textit{WMAP}9+
data. In this paper, we will have a close look at the constraints
on the brane inflation models with the results from
\textit{Planck} \cite{Planck13-1,Planck13-16,Planck13-22} and
pre-\textit{Planck} surveys
\cite{Hinshaw12,Hou12,Sievers13,Anderson13}.

This paper is organized as follows: In Sec.~\ref{sec:efolds}, we
discuss the relationship between the $e$-folding number of
inflation and the pivot scale of observation. In
Sec.~\ref{sec:Dp}, we discuss a simple brane inflation model
neglecting the problem of dynamic stabilization. This is the
simplest brane inflation model one can achieve in the
multidimensional spacetime. In Sec.~\ref{sec:kklmmt}, we focus on
the KKLMMT model and compare the model predictions with
observational data. In Sec.~\ref{sec:dbi}, we turn to the
discussion of the infrared DBI inflation model and confront the
model predictions with observational data. The conclusion is
presented in the last section.

\section{Number of \textit{e}-folds}
\label{sec:efolds} Before we start to constrain any inflation
model, we first address an important issue in the inflation model
tests: how do we compare model predictions with observational
data? Inflation models are actually models of different inflation
potentials, where the amplitude and shape are the features of
various models. In the community of inflation theorists, people
use the amplitude of potential to characterize the energy scale of
inflation and a set of ``slow-roll'' parameters to describe the
shape of the potential.
Usually the shape of the potential includes the ``slope'' and
``curvature'' parameters of the potential.
For a given potential, the slow-roll parameters can be expressed in
terms of the number of \textit{e}-folds ($N_e$) which characterizes
the duration of inflation.

On the other hand, observations from the CMB provide constraints
on the amplitude and shape of the primordial power spectrum. But
since the power spectrum itself is a $k-$dependent quantity, the
measured amplitude ($\Delta^{2}_{\cal R}$), tilt ($n_{s}$),
tensor-to-scalar ratio ($r$) and running of spectral index
($dn_{s}/d \ln k$) are referred to a particular ``pivot scale''.
This indicates that for a given data set, if the pivot scale is
switched to a different value, the constraints on the
($\Delta^{2}_{\cal R}(k_{0})$, $n_{s}(k_{0})$, $d n_{s}/d \ln k$)
can be slightly different. Therefore, to really compare model
predictions with observational data, we need to associate the
number of \textit{e}-folds with the pivot scale of observation.
Our main goal in this section is to obtain a relationship between
the number of \textit{e}-folds $N_{e}$ and its corresponding
comoving scale $k$.

Once inflation happened, different scales (different $k-$modes)
stretched out of the Hubble radius at different time. After
inflation, the Universe experienced a short period of reheating, and
then entered into radiation, matter and dark energy dominated eras.
The number of \textit{e}-folds is related to the processes of
subsequent evolution because both the inflation and subsequent
evolutionary processes contribute to the total expansion factor of
the Universe (see Fig.~1 in \cite{Liddle03}). We can therefore write
\cite{Liddle03}
\begin{eqnarray}
\frac{k}{a_{0}H_{0}} = \frac{a_{\rm{k}}H_{\rm{k}}}{a_{0}H_{0}} =
\left(\frac{a_{\rm{k}}}{a_{\rm{e}}}
\right)\left(\frac{a_{\rm{e}}}{a_{\rm{reh}}}\right)
\left(\frac{a_{\rm{reh}}}{a_{\rm{eq}}} \right)
\left(\frac{H_{\rm{k}}}{H_{\rm{eq}}}\right)
\left(\frac{a_{\rm{eq}}H_{\rm{eq}}}{a_{0}H_{0}}\right),
\label{nk1}
\end{eqnarray}
where we used the subscripts ``$\rm{k}$, $\rm{e}$, $\rm{reh}$,
$\rm{eq}$, $0$'' to represent the horizon exit, end of inflation,
reheating epoch, matter-radiation equality epoch and present time.
Number of \textit{e}-folds between horizon exit and the end of
inflation is $N_{e}(k)=\ln (a_{e}/a_{\rm k})$. By assuming the
equation of state during the reheating era being $w$ ($w=P/\rho$),
one can reach the following equation (see also
\cite{Liddle00,Liddle03}),
\begin{eqnarray}
N_{e}(k)&=& -\ln \left(\frac{k}{a_{0}H_{0}} \right)+ \ln
\left(\sqrt{\frac{V_{\rm{k}}}{3
M^{2}_{\textrm{pl}}}}\frac{1}{H_{\rm{eq}}}
\right) +\ln (219\Omega_{\rm{m}}h) \nonumber \\
&& -\frac{1}{3(1+w)}\ln
\left(\frac{\rho_{\rm{e}}}{\rho_{\rm{reh}}} \right)-\frac{1}{4}\ln
\left(\frac{\rho_{\rm{reh}}}{\rho_{\rm{eq}}} \right) ,\label{nk2}
\end{eqnarray}
where $V_{\rm{k}}$ is the energy scale of inflation at horizon exit,
and $M_{\textrm{pl}}\equiv 1/(8\pi G) \simeq 2.4 \times 10^{18}
\textrm{GeV}$ is the reduced Planck mass. By defining the ratio of
the energy densities between at the reheating and at the end of
inflation as $x\equiv \rho_{\rm{reh}}/\rho_{\rm{e}}$, and regarding
$\rho_{\rm{e}}=V_{\rm{k}}$ (``slow-roll'' approximation), we can
rewrite Eq.~(\ref{nk2}) as
\begin{eqnarray}
N_{e}(k)&=& -\ln \left(\frac{k}{a_{0}H_{0}} \right)+ \ln
\left(\sqrt{\frac{V_{\rm{k}}}{3
M^{2}_{\textrm{pl}}}}\frac{1}{H_{\rm{eq}}}
\right) +\ln (219\Omega_{\rm{m}}h) \nonumber \\
&& + \left(\frac{1}{3(1+w)}-\frac{1}{4} \right)\ln x+
\frac{1}{4}\ln \left(\frac{\rho_{\rm{eq}}}{V_{\rm{k}}} \right)
.\label{nk3}
\end{eqnarray}
To further simplify this equation, we use the requirement that the
primordial perturbations have to produce the observed level of
fluctuations ($P_{s}(k_{0})\simeq 2.43 \times 10^{-9}$), i.e.,
\begin{eqnarray}
P_s=\frac{V_{\rm{k}}/M_{\rm pl}^4}{24 \pi^{2} \epsilon_{v}}, \textrm{where   }
\epsilon_{v}=\frac{M^{2}_{\textrm{pl}}}{2}\left(\frac{V'}{V}
\right)^{2}. \label{ps1}
\end{eqnarray}
Substituting known quantities, Eq.~(\ref{nk3}) can be greatly
simplified as
\begin{eqnarray}
N_{e}(k) &=& -\ln \left(\frac{k}{2.33\times
10^{-4}\textrm{Mpc}^{-1}}\right) +63.3+\frac{1}{4}\epsilon_{v}
\nonumber \\ &&+ \left(\frac{1}{3(1+w)}-\frac{1}{4} \right)\ln
x.\label{nk4}
\end{eqnarray}
For a particular mode $k$, its corresponding $N_{e}(k)$ relies on
the equation of state $w$ and energy scale of reheating
$\rho_{\rm{reh}}$. Since the standard picture tells that vacuum is
decayed into standard particles, $\rho_{\rm{reh}}$ is always less
than or equal to potential energy scale $V_{k}$, i.e. $x\leq1$, thus
$\ln x$ is always a negative value. Therefore, if $w\rightarrow 0$
(close to a ``matter-dominated phase''), the fourth term of
Eq.~(\ref{nk4}) becomes $(1/12)\ln x$, which gives a minimal number
of \textit{e}-folds. This means that if the equation of state is
close to zero, the shape of the potential ($\sim \phi^{2}$) can keep
inflaton oscillating for a fairly long period of time while the
Universe is expanding, therefore we need less number of
\textit{e}-folds to produce an observable scale of the Universe. On
the other hand, if the equation of state during the reheating era is
$w \simeq 1/3$, or the reheating is instantaneous
($\rho_{\rm{reh}}=V_{\rm{k}}$, i.e., $\ln x=0$), the fourth term
vanishes, which gives the maximum number of \textit{e}-folds ($\sim
\phi^{4}$). Since there is a great uncertainty of what energy scale
reheating really happened, in the following discussion we stick to
the case of instantaneous reheating, so that the number of
\textit{e}-folds becomes
\begin{eqnarray}
N_{e}(k) = -\ln \left(\frac{k}{2.33\times
10^{-4}\textrm{Mpc}^{-1}}\right) +63.3+\frac{1}{4}\epsilon_{v}
\label{nk5}.
\end{eqnarray}
For joint \textit{WMAP}9+SPT+ACT+BAO+$H_{0}$ data ($k_{0}=0.002
\textrm{Mpc}^{-1}$) and \textit{Planck}+WP data ($k_{0}=0.05
\textrm{Mpc}^{-1}$), the corresponding numbers of \textit{e}-folds
are
\begin{eqnarray}
N(k_{0})=61.2 + \frac{1}{4} \ln \epsilon  ~~~\textrm{(for~\textit{WMAP}9+)}, \nonumber \\
N(k_{0})=58.2 + \frac{1}{4} \ln \epsilon  ~~~
\textrm{(for~}{\textit{Planck}+{\rm WP})}. \label{pivot}
\end{eqnarray}

Typically observational predictions of slow-roll parameters (e.g.
$\epsilon_{v}$) depend on $N_{e}$, so both sides of
Eq.~(\ref{pivot}) contain $N_{e}$ which could be solved
simultaneously. In practice, the deviation of $N_{e}$ from the
typical value $60$ is always small, so one can solve
Eq.~(\ref{pivot}) iteratively by assuming a particular $N_e$ and use
it to calculate the potential properties, then use these to
recalculate $N_e$, and so on. In fact, one iteration easily suffices
to give sufficient accuracy of $N_{e}$. We will illustrate this in
the following sections.

\section{A toy model}
\label{sec:Dp}
\subsection{Model predictions}
To begin with, we consider a toy model of brane inflation; actually,
this is a prototype of the brane inflation: a pair of $Dp$ and
$\bar{D}p$-branes ($p\geq 3$) fill the four large dimensions and are
separated from each other in the extra six dimensions that are
compactified. Note that this model is not a realistic working model
because it does not take into account the warped space-time and
moduli stabilization. However, such a prototype provides us with a
warm-up exercise for comparing models with CMB observations. In this
model, the inflaton potential is given by
\cite{Quevedo02,Huang06,Ma09}
\begin{equation}
V=V_{0}\left(1-\frac{\mu^n }{\phi^n }\right),  \label{potential1}
\end{equation}
where $V_0$ is an effective cosmological constant on the brane and
the second term in Eq.~(\ref{potential1}) is the attractive force
between the branes. The parameter $n$ has to satisfy $n\leq 4$
because the transverse dimension has to be less or equal to $6$. The
$e$-folding number $N_{e}$ at the horizon exit before the end of
inflation is related to the field value as \cite{Huang06,Ma09}
\begin{equation}
\phi _{N}=[N_{e} M_{\rm{pl}}^{2}\mu ^{n}n(n+2)]^{1/(n+2)}.
\label{begin1}
\end{equation}
The slow-roll parameters have been calculated as
\cite{Quevedo02,Huang06,Ma09}
\begin{eqnarray}
\epsilon _{v} &=&
\frac{M_{\rm{pl}}^{2}}{2}\left(\frac{V'}{V}\right)^{2} \nonumber
\\
&=&\frac{n^{2} }{2(n(n+2))^{\frac{2(n+1)}{n+2}}}\left(\frac{\mu
}{M_{\rm{pl}}}\right)^{\frac{2n}{n+2}}N_{e}^{-
\frac{2(n+1)}{n+2}}, \label{epsilon1}
\end{eqnarray}%
\begin{equation}
\eta
_{v}=M_{\rm{pl}}^{2}\frac{V''}{V}=-\frac{n+1}{n+2}\frac{1}{N_{e}},
\label{eta1}
\end{equation}%
\begin{equation}
\xi
_{v}=M_{\rm{pl}}^{4}\frac{V'V'''}{V^{2}}=\frac{n+1}{n+2}\frac{1}{N_{e}^{2}}.
\label{xi1}
\end{equation}%
The observational quantities, $n_{s}$, $r$, and $\alpha_s$ (spectral
index, tensor-to-scalar ratio, and running of spectral index), can
be expressed as the combination of slow-roll parameters
\begin{eqnarray}
n_{s} &=& 1+2 \eta_{v}-6 \epsilon_{v}, \nonumber \\
r &=& 16 \epsilon_{v}, \nonumber \\
\alpha_{s} &=& -24 \epsilon^{2}_{v} +16
\epsilon_{v}\eta_{v}-2\xi_{v}. \label{power1}
\end{eqnarray}%
These are the observables that we will compare with observational
results.

\subsection{Constraints from \textit{Planck} and pre-\textit{Planck} data}
\label{sec:model1compare}

\begin{figure}[tbh]
\centerline{\includegraphics[bb=0 0 874 447,
width=3.6in]{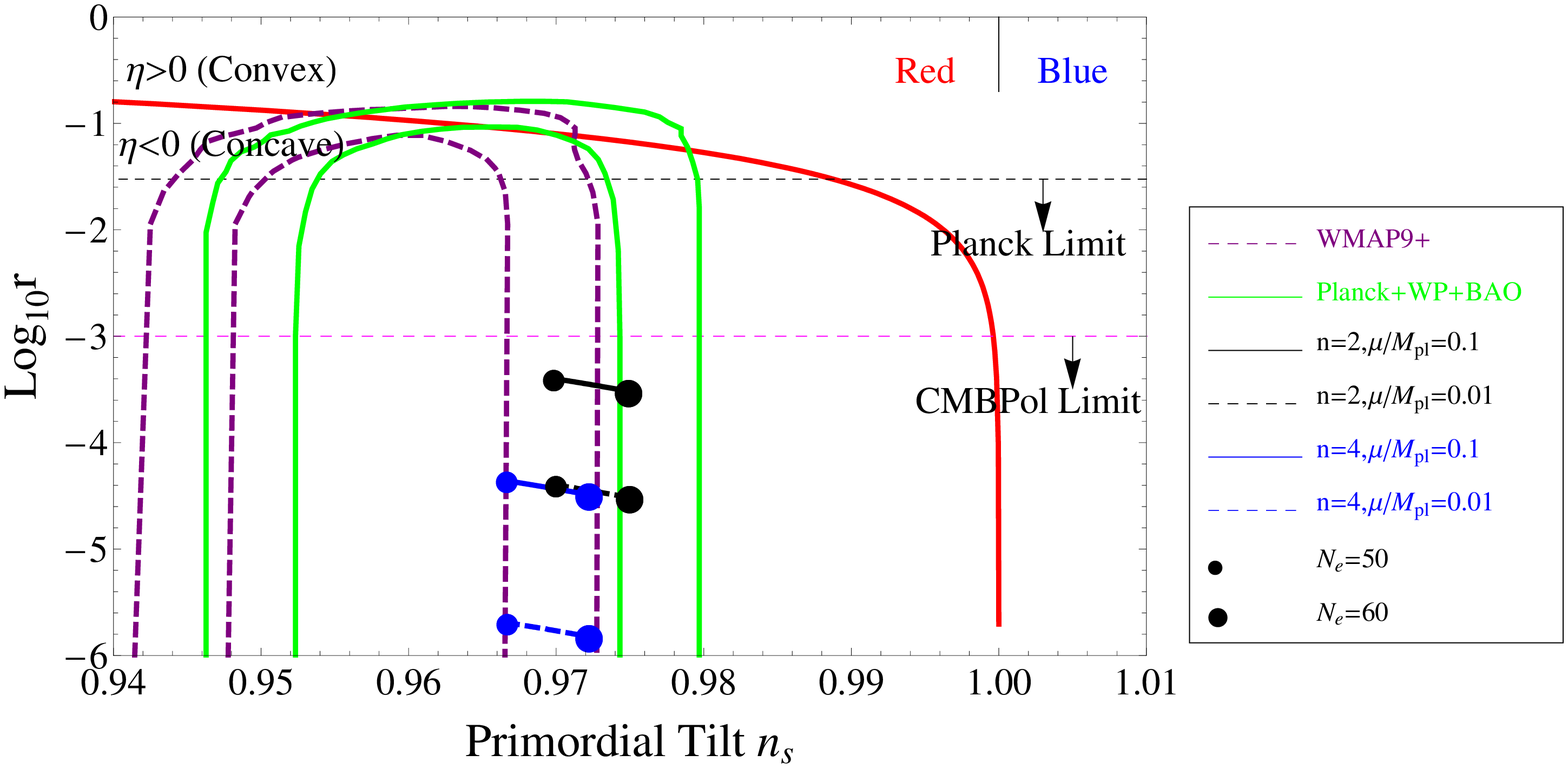}} \caption{The $r$--$n_{s}$ plot for
theoretical models, current observational constraints and
predicted limits from \textit{Planck} polarization maps and
CMBPol. For model predictions: the red curve across the whole
diagram is the divided line for $\eta=0$, on either side the
potential has different curvatures as marked onto the plot. The
black and blue lines are the predictions for the $n=2$ and $n=4$
models with $\mu/M_{\rm pl}=0.1$ (solid line) and $0.01$ (dashed
line). The small and big dots correspond to $N_{e}=50$ and 60,
respectively. We also mark the red tilt and blue tilt on the top
of the diagram. For the observational results: the purple dashed
contours are the joint constraints from
\textit{WMAP}9+SPT+ACT+BAO+$H_{0}$ (``\textit{WMAP}9+'') and the
green solid contours are the joint constraints from
\textit{Planck}+WP+BAO. The two horizontal dashed lines are the
predicted observational limits of tensor-to-scalar ratio $r$ from
\textit{Planck} polarization map ($r \lesssim 0.03$)
\cite{Efstathiou09} and CMBPol ($r \lesssim 0.001$)
\cite{Baumann09,Ma10}; note that the two lines are not from actual
data, but are based on the predictions of future data. }
\label{fig1:model1rns}
\end{figure}

\begin{figure*}[tbh]
\centerline{\includegraphics[bb=0 0 737 459,
width=3.3in]{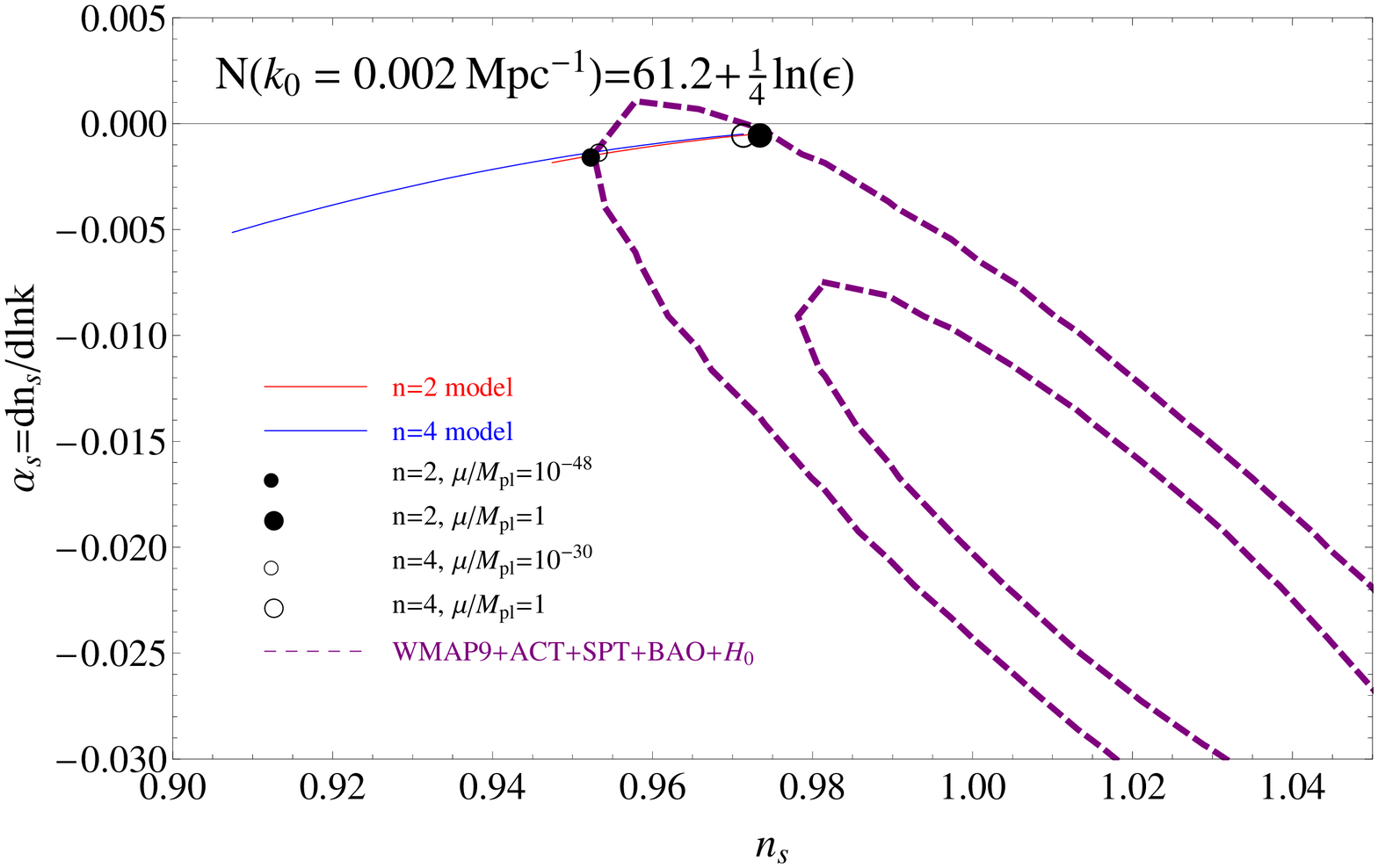}
\includegraphics[bb=0 0 667 421,width=3.3in]{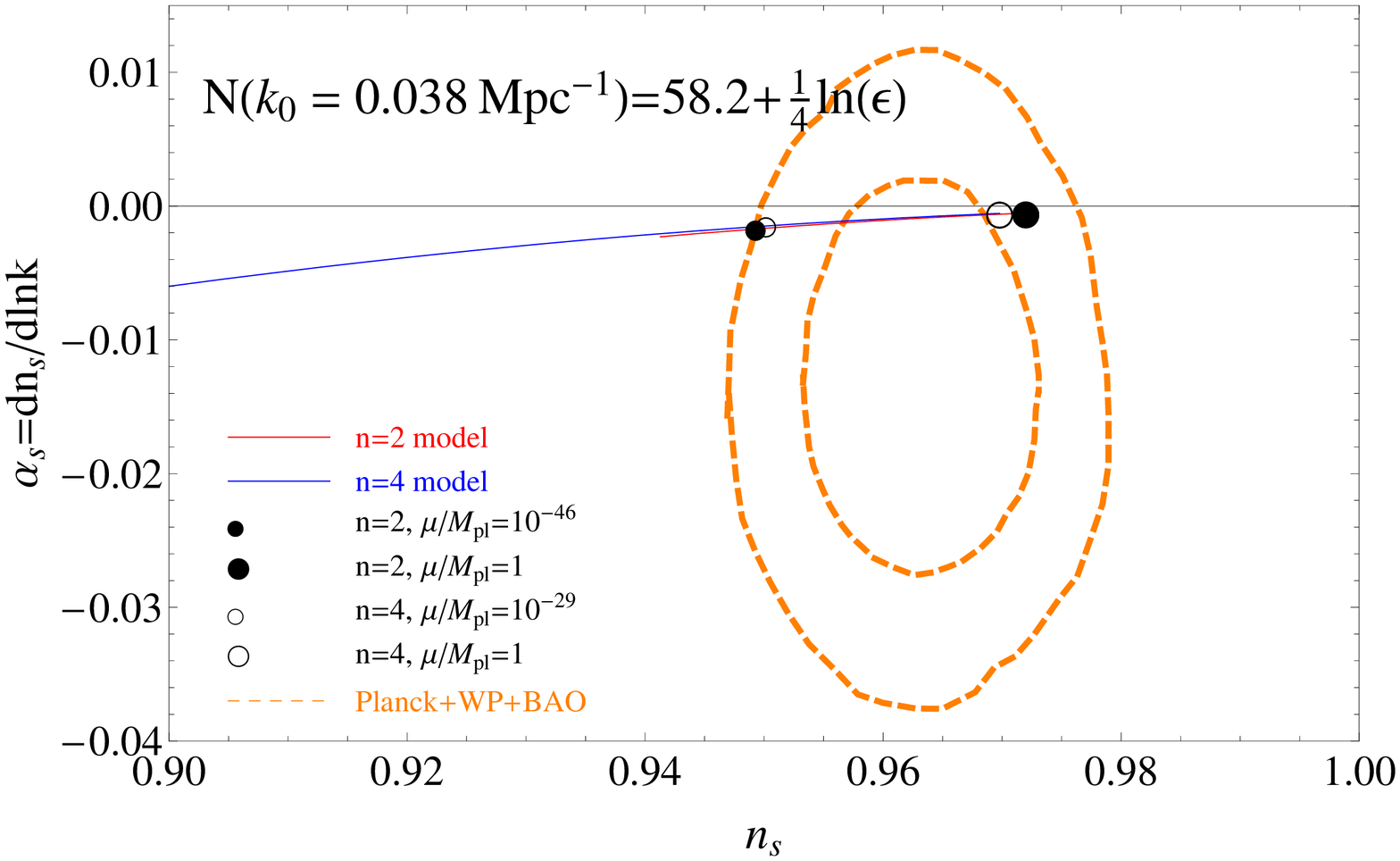}}
\caption{Comparing the prediction of the prototype of brane
inflation with the observational constraints on the $d n_{s}/d \ln
k$--$n_s$ plane. \textit{Left}-- Comparing with the joint
constraints from combination of \textit{WMAP} 9-year data, ACT,
SPT, BAO and $H_{0}$ data at the pivot scale $k_0=0.002$
{Mpc}$^{-1}$. \textit{Right}-- Comparing with the joint
constraints from \textit{Planck}+WP+BAO data at the pivot scale
$k_{0}=0.038$ {Mpc}$^{-1}$. In both panels, the number of
\textit{e}-folds of model predictions matches the pivot scale of
the constrained contours. See text for more details of the
theoretical predictions of the model.} \label{fig1:model1asns}
\end{figure*}

In Fig.~\ref{fig1:model1rns}, we plot the theoretical prediction
of $r$ (in terms of $\log_{10}r$) and $n_{s}$. The black and blue
lines are the predictions for $n=2$ and $n=4$ models with
$\mu/M_{\rm pl}$ values being $0.1$ and $0.01$. The range between
small and big dots corresponds to the number of \textit{e}-folds
within $[50,~60]$. The red line across the diagram is the boundary
line between the convex potential ($\eta_{v}>0$) and concave
potential ($\eta_{v}<0$). We also plot the $1\sigma$ and $2\sigma$
constraints on $r$ and $n_{s}$ from \textit{WMAP}9+ data and
\textit{Planck}+WP+BAO (hereafter \textit{Planck}+) data. From the
plot, one can see that \textit{WMAP}9+ prefers a slightly lower
$n_{s}$ comparing with \textit{Planck}+ data. In addition, most of
the contour regions locate within $\eta_{v}<0$ region, indicating
strong evidence of concave potential. The models with $n=2$ and
$n=4$ lying within the contours suggests that the model prediction
is consistent with the current constraints. We also plot the
predicted detection limits of $r$ from \textit{Planck}
polarization experiment \cite{Efstathiou09} and CMBPol
\cite{Baumann09}; note that these two limits are not from actual
data, but are based on the predictions of future data. One can see
that even if $\mu/M_{\rm{pl}}$ is of order $0.1$, the model
prediction is still much lower than the CMBPol detection limit.
Only if $\mu/M_{\rm{pl}}>0.3$, could the CMBPol be able to detect
the tensor mode in this model.

In Fig.~\ref{fig1:model1asns} we plot the predicted
$\alpha_{s}$--$n_{s}$ relation for the brane inflation model with
the constraint results from \textit{WMAP}9+ and \textit{Planck}+
data. The purple contours on the left panel is the joint
constraints on $\alpha_{s}$--$n_{s}$ from the \textit{WMAP}9+ data
set with the pivot scale $k_0=0.002~\textrm{Mpc}^{-1}$. Therefore
we use Eq.~(\ref{pivot}) to determine the number of
\textit{e}-folds: we substitute a fiducial number of
\textit{e}-folds $N_{\rm{fid}}=60$ into Eq.~(\ref{epsilon1}) and
obtain an estimate of $\epsilon$, then substitute it into
Eq.~(\ref{pivot}) to obtain the corresponding number of
\textit{e}-folds for this model. We test that one iteration is
enough for determine the specific $N_{e}$. Then  with
Eqs.~(\ref{epsilon1})--(\ref{power1}) we plot the
$\alpha_{s}$--$n_{s}$ prediction with variation of the parameter
$\mu$. The red line is for the $n=2$ model and the blue line is
for the $n=4$ model. The two lines are pretty close to each other,
and they are all outside $1\sigma$ confidence level (CL) but some
range is within $2\sigma$ CL. We then figure out which values of
$\mu$ can match the results inside the $2\sigma$. We give a couple
of trials and find that, for the $n=2$ model $\mu/M_{\rm{pl}}$
needs to be between $10^{-48}$ and unity, and for the $n=4$ model
this range is $[10^{-30},~1]$. On the right panel of
Fig.~\ref{fig1:model1asns}, we use the constraints from
\textit{Planck}+WP+BAO to compare with the theoretical
predictions. The results are similar to the left panel, except
that the range of $\mu/M_{\rm{pl}}$ is shorten to be
$[10^{-46},~1]$ for the $n=2$ model, and $[10^{-29},~1]$ for the
$n=4$ model. In a word, the prototype of brane inflation with
potential form (\ref{potential1}) is consistent with the
observational constraints on $\alpha_{s}$ and $n_{s}$.

Then let us see what this implies for the energy scale of
inflation in this model. The amplitude of the scalar perturbations
is \cite{Firouzjahi05,Huang06,Liddle00}
\begin{equation}
\Delta^{2}_{\mathcal{R}}=\frac{V}{M^{4}_{\rm{pl}}}
\frac{1}{24\pi^2 \epsilon_{v}}, \label{eq:amplitude}
\end{equation}
which is constrained to be $\sim 2.2\times 10^{-9}$ by the
\textit{Planck} data \cite{Planck13-1}. We substitute $\epsilon_{v}$
[Eq.~(\ref{epsilon1})] into Eq.~(\ref{eq:amplitude}), and thus we
obtain a relationship between the amplitude of inflation and the
parameter $\mu$,
\begin{equation}
\frac{V^{\frac{1}{4}}}{M_{\rm{pl}}}=\left(24 \pi^2
\frac{n^{2}}{2(n(n+2))^{\frac{2(n+1)}{n+2}}}
\left(\frac{\mu}{M_{\rm{pl}}} \right)^{\frac{2n}{n+2}}
N^{-\frac{2(n+1)}{n+2}}_{e} \right)^{\frac{1}{4}}.
\label{eq:amplitude1}
\end{equation}
Then from our estimation of $\mu$ we can find that the amplitude of
inflation is in the range $[2.7 \times 10^{4},~ 8.4 \times
10^{15}]~\textrm{GeV}$ for the $n=2$ model and  $[1.3 \times 10^{6},
~5.9 \times 10^{15}]~\textrm{GeV}$ for the $n=4$ model. These are
all reasonable ranges for $V$, because it needs to be lower than
$10^{16} \textrm{GeV}$ so that we do not detect any tensor mode yet,
and greater than the particle physics energy scale $10^{3}
\textrm{GeV}$ since the inflaton is not detected in LHC.

\section{KKLMMT model}
\label{sec:kklmmt}

The prototype of the brane inflation model discussed above is not
a realistic model, because the distance between the brane and the
antibrane would be larger than the size of the extra-dimensional
space if the inflaton is slowly rolling in this scenario
\cite{Huang06,Ma09}. It indicates that this model is not really
reliable from the viewpoint of theory itself. The first more
realistic brane inflation model which considers the effect of
warped spacetime on inflaton potential is the so-called KKLMMT
model \cite{Kachru03}, whose predictions are directly calculable
and can be directly compared to observations. Note that, strictly
speaking, the KKLMMT model is only a brane-inflation-inspired
model rather than a scenario with all elements of the potential
computed precisely; for more complicated versions of brane
inflation, see \cite{Baumann08,Baumann10}.

\subsection{Model predictions}
\label{sec:kklmmtpredict} The KKLMMT model is derived from the
type IIB string theory. In the model, the spacetime contains
highly warped compactifications, and all moduli stabilized by the
combination of fluxes and nonperturbative effects
\cite{Kachru03,Ma09}. Once a small number of $\overline{\rm
D}$3-branes are added, the vacuum can be successfully lifted to de
Sitter state. Furthermore, one can add an extra pair of D3-brane
and $\overline{\rm D}$3-brane in a warped throat with the D3-brane
moving towards the $\overline{\rm D}$3-brane that is located at
the bottom of the throat. When the $\rm{D}$3 moves towards the
$\overline{\rm D}$3, inflation takes place; therefore, the
scenario of brane inflation can be achieved in this model. The
warped throat successfully guarantees a flat potential, which
solves the ``$\eta$ problem'' in the brane inflation.

Let us start with the inner space of the Calabi-Yau manifold,
where the geometry is highly warped and its spacetime can be
approximate $AdS_{5}\times X_{5}$ form. The $AdS_{5}$ metric in
Poincar\'{e} coordinates has the form
\cite{Ma09,Firouzjahi05,Baumann06}
\begin{equation}
ds^{2}=h^{-\frac{1}{2}}(r)(-dt^2+a(t)^2d\vec{x}^2)+h^{\frac{1}{2}
}(r)ds_6^2,  \label{warp metric}
\end{equation}
where $h(r)$ is the warp factor,
\begin{equation}
h(r)=\frac{R^4}{r^4},  \label{warped factor}
\end{equation}%
where we express the radius of curvature of the $AdS_{5}$ throat
as $R$. The potential within the warped throat is
\begin{equation}
V(\phi )=\frac{1}{2}\beta H^{2}\phi ^{2}+2T_{3}h^{4}(1-\frac{\mu
^{4}}{\phi ^{4}}),  \label{V}
\end{equation}%
which basically constitutes three terms. The first term is the
K\"{a}hler potential term which arises from interactions of
superpotentials \cite{Firouzjahi05} where $H$ is the Hubble
parameter and $\beta$ describes the coupling between inflaton $\phi$
(position of D3 brane) and space expansion. In general, the value of
$\beta$ depends on $\phi$ value because the conformal coupling
depends on the position of the D3 brane, but we expect that $\beta$
to stay more or less constant in each throat, so approximately
$\beta \simeq \rm{const}$ here \cite{Firouzjahi05}. Generically
$\beta \sim 1$, but for KKLMMT type of slow-roll model, $|\beta |$
is much less than unity. The second term ($2T_{3}h^{4}$) is the
effective cosmological constant
in the brane~\cite{Firouzjahi05}. This is the term that drives the
accelerated expansion of the Universe. The last term (with minus
sign) provides the Coulomb-like attractive potential between the
D3-brane and the $\bar{\rm{D}}$3-brane, making the two branes
eventually collide. Note that $T_{3}$ is the D3-brane tension and
it is related to $\mu$ through $\mu ^{4}=\frac{27}{32\pi
^{2}}T_{3}h^{4}$. We then have
\begin{equation}
V(\phi )=\frac{1}{2}\beta H^{2}\phi ^{2}+\frac{64\pi ^{2}\mu ^{4}}{27}(1-%
\frac{\mu ^{4}}{\phi ^{4}}).  \label{V1}
\end{equation}%
Under the slow roll approximation, the Friedmann equation becomes
\begin{equation}
3M_{\rm{pl}}^{2}H^{2}\simeq V(\phi )\simeq V_{0}=\frac{64\pi
^{2}\mu ^{4}}{27}; \label{Friedman}
\end{equation}%
therefore, $\mu$ also represents the energy scale of inflation.

Given the potential, it becomes a standard calculation to obtain
the field value at the onset of inflation and the set of slow-roll
parameters. Following \cite{Firouzjahi05,Huang06,Ma09}, we have
\begin{equation}
\phi _{N}^{6}=24M_{\rm{pl}}^{2}\mu ^{4}m(\beta ),  \label{ending2}
\end{equation}
where
\begin{equation}
m(\beta )=\frac{e^{2\beta N}(1+2\beta )-(1+\frac{1}{3}\beta )}{2\beta (1+%
\frac{1}{3}\beta )}.  \label{m(beta)}
\end{equation}

Therefore, the slow-roll parameters in the KKLMMT model are
\begin{equation}
\epsilon _{v}=\frac{1}{18}\left(\frac{\phi
_{N}}{M_{\rm{pl}}}\right)^{2}\left[\beta +\frac{1}{2m(\beta
)}\right]^{2}, \label{epsilon2}
\end{equation}
\begin{equation}
\eta_v =\frac{\beta }{3}-\frac{5}{6}\frac{1}{m(\beta )},
\label{yita3}
\end{equation}
\begin{equation}
\xi _{v}=\frac{5}{3}\frac{1}{m(\beta )}\left[\beta
+\frac{1}{2m(\beta )}\right]. \label{xi2}
\end{equation}

Now we need to use the observed CMB fluctuations to fix the
amplitude of the scalar perturbations. Similar to the calculation we
did in Sec.~\ref{sec:model1compare}, we obtain
\begin{equation}
\Delta _{\cal R}^{2}\simeq \frac{V}{M_{\rm{pl}}^{4}}\frac{1}{24\pi
^{2}\epsilon _{v}}=\frac{2}{27m(\beta )}\left(\beta
+\frac{1}{2m(\beta )}\right)^{-2}\left(\frac{\phi
_{N}}{M_{\rm{pl}}}\right)^{4}, \label{amplitude}
\end{equation}
and thus we have
\begin{equation}
\epsilon
_{v}=\frac{1}{48}\left(\frac{3}{2}\right)^{\frac{1}{2}}(\Delta
_{\cal R}^{2})^{\frac{1}{2}}m(\beta )^{-\frac{5}{2}}(1+2\beta
m(\beta ))^{3}. \label{epsilon3}
\end{equation}
The \textit{Planck} data give the amplitude of the primordial
scalar power spectrum as $\Delta _{\cal R}^{2}\simeq 2.2\times
10^{-9}$ for $N\sim 50$ \cite{Planck13-16}. Therefore, all of the
slow-roll parameters in the KKLMMT model
[Eqs.~(\ref{epsilon2})--(\ref{xi2})] are related to the parameter
$\beta$ and the number of \textit{e}-folds $N_{e}$. Following
Eq.~(\ref{power1}), we will use parameters $n_{s}$, $\alpha_{s}$
and $r$ to figure out the best $\beta$ value given the current
observational data.

\subsection{Constraints from observational data}

\begin{figure*}[tbh]
\centerline{\includegraphics[bb=100 0 665
432,width=3.5in]{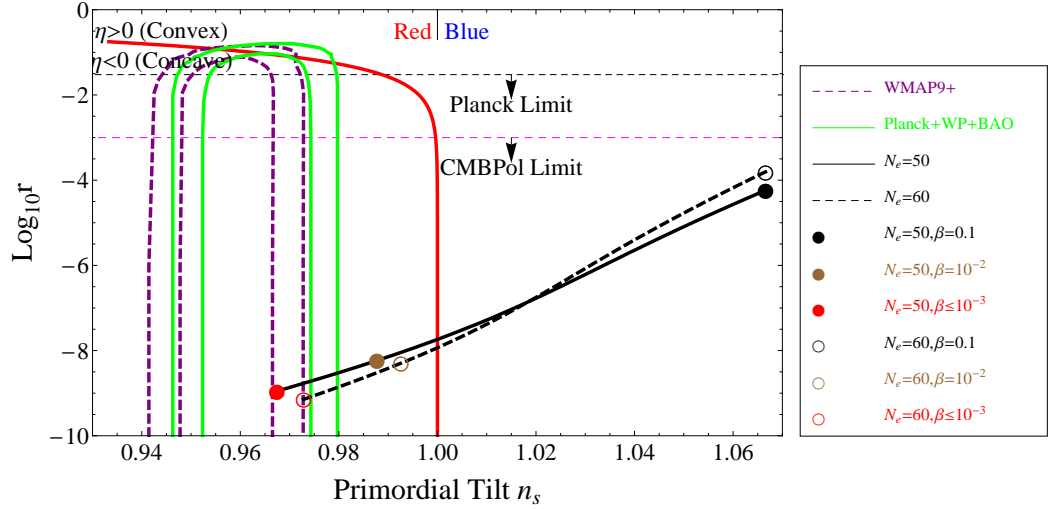}} \caption{Similar plot as
Fig.~\ref{fig1:model1rns} but for the KKLMMT model. The black solid
and black dashed lines represent the trajectories for $N_{e}=50$ and
$60$, respectively. The empty and filled circles mark the points
where the model takes $\beta=0.1$ (black), $0.01$ (brown) and $\leq
0.001$ (red), respectively.} \label{fig1:model2rns}
\end{figure*}

\begin{figure}[tbh]
\centerline{\includegraphics[bb=0 0 809 511,
width=3.5in]{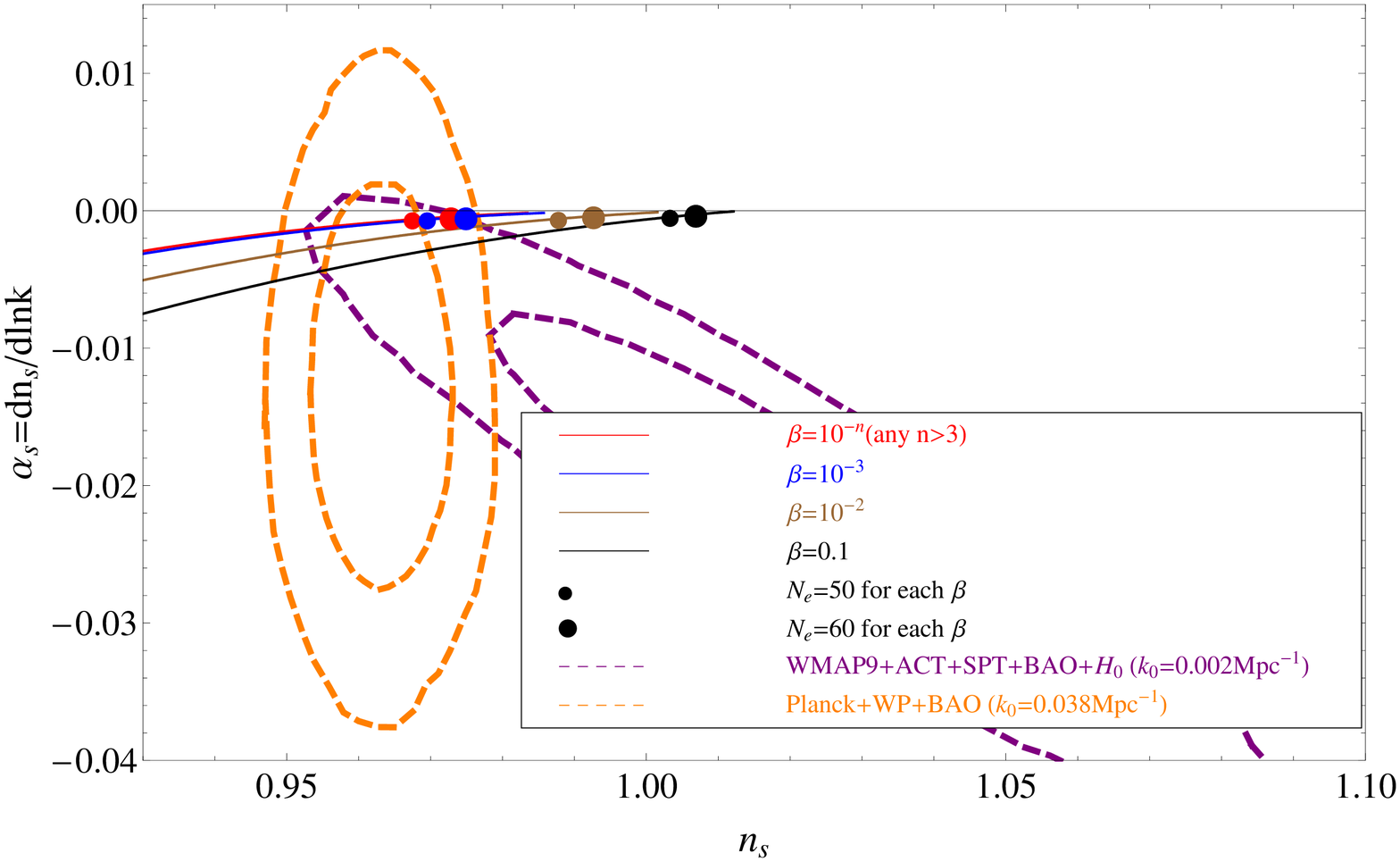}} \caption{Comparison of the joint
observational constraints with the KKLMMT model predictions on the
$\alpha_{s}$--$n_{s}$ plane. The purple and orange contours are
the results from \textit{WMAP}9+ and \textit{Planck}+WP+BAO. The
black, brown and blue lines are for models with $\beta=0.1,
10^{-2}$, and $10^{-3}$, respectively. The red line is for any
model with $\beta<10^{-3}$. The small and big color dots denote
$N_{e}=50$ and $60$, respectively.} \label{fig1:model2}
\end{figure}

In Fig.~\ref{fig1:model2rns}, we plot the $r$--$n_{s}$ diagram
similar to the structure of Fig.~\ref{fig1:model1rns}. Instead,
here it is the KKLMMT model. The black solid and black dashed
lines represent the trajectories for $N_{e}=50$ and $60$,
respectively. Different colors of empty and filled circles mark
the point where the model takes different $\beta$ values. One can
see how the $\beta$ parameter controls the shape of the potential.
If it is greater than $0.03$, the potential turns to be convex
which is not preferred by current observational data. Actually,
the problem for $\beta>0.01$ is that it provides a blue tilt which
has already been ruled out by \textit{Planck}+WP+BAO at more than
$5\sigma$ CL. In order for the model to pass this test, $\beta$
value has to be much smaller than $10^{-3}$. In fact, since the
current \textit{Planck} data prefer the $n_{s}$ value around
$0.96$ (green contours), the models with $\beta<10^{-3}$ are just
about to survive since they offer the spectral index to be $0.96$
but not smaller than $0.95$ (see Fig.~\ref{fig1:model2} as well).
This means that as long as the CMB data prefer $n_{s}$ to be
around $0.96$, this model can always pass this test and survive.
Nevertheless, the parameter needs to be highly fine-tuned.
Finally, similar to Fig.~\ref{fig1:model1rns}, one can see that
the tensor mode predicted by the KKLMMT model is really
undetectably small since it is several orders of magnitude lower
than the \textit{Planck} polarization \cite{Efstathiou09} and
CMBPol limits \cite{Baumann09,Ma09}.

In Fig.~\ref{fig1:model2} we show the comparison of the
observational constraints and the model predictions on the
$\alpha_{s}$--$n_{s}$ plane. One can see that
\textit{Planck}+WP+BAO prefers a slightly negative running with a
very red power tilt. The tilt of the power spectrum at more than
$5\sigma$ deviates from unity (the Harrison-Zel'dovich spectrum),
while the running is at less than $2\sigma$ away from zero. On the
other hand, if $\alpha_{s}$ is released as a free parameter, the
\textit{WMAP}9+ data set cannot tighten up $n_{s}$ to be less than
unity. The purple contours stretch from a small negative running
($\sim -0.01$) with red tilt ($\sim 0.96$) out to a large negative
running ($\sim -0.04$) with blue tilt ($\sim 1.05$) region.
However, constraints from these two different data sets overlap at
the small negative running and red tilt region, indicating that
this is the preferable region for both data sets. In addition, we
plot the model predictions for different $\beta$ values, and we
mark the region of model predictions in between $N_{e}=50$ and
$60$ in order to have a direct vision of whether this ``physically
plausible'' region falls in the observational constraint contours.
One can also see that the KKLMMT model cannot produce a red tilt
and suitable level of negative running unless $\beta \leq 10^{-3}$
at $2\sigma$ CL. The model with $\beta=0.01$ cannot fit the
$2\sigma$ joint constraints in either case. This is actually an
order of magnitude tighter than the previous upper limit of
$\beta$ from \textit{WMAP} 5-year data \cite{Ma09} ($\beta<0.01$
at $2\sigma$ CL), and also much tighter than the combined
constraints ($\beta<6 \times 10^{-3}$) from
\textit{WMAP}3+SDSS~\cite{Huang06}.

\section{IR DBI model}
\label{sec:dbi}

\begin{figure}[tbh]
\centerline{\includegraphics[bb=0 0 665 419,
width=3.3in]{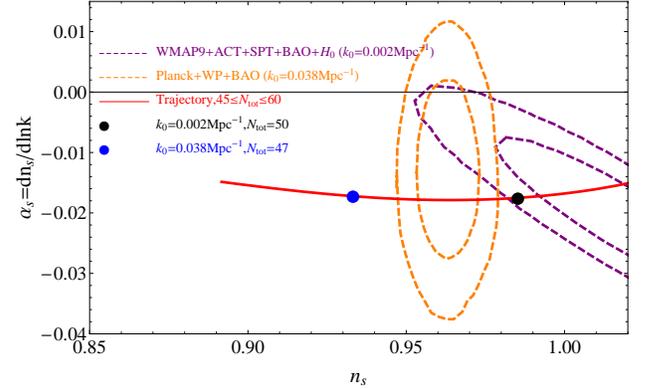}} \caption{Comparison of the constraints
on $\alpha_{s}$--$n_{s}$ with the IR DBI model predictions. The red
line is the trajectory of the number of \textit{e}-folds in range of
$45$--$60$. The blue and black dots corresponds to the particular
numbers of \textit{e}-folds for the pivot scales $k_0=0.038
\textrm{Mpc}^{-1}$ and $k_0=0.002 \textrm{Mpc}^{-1}$.}
\label{fig1:model3asns}
\end{figure}

\begin{figure*}[tbh]
\centerline{\includegraphics[bb=0 0 645 413,
width=3.45in]{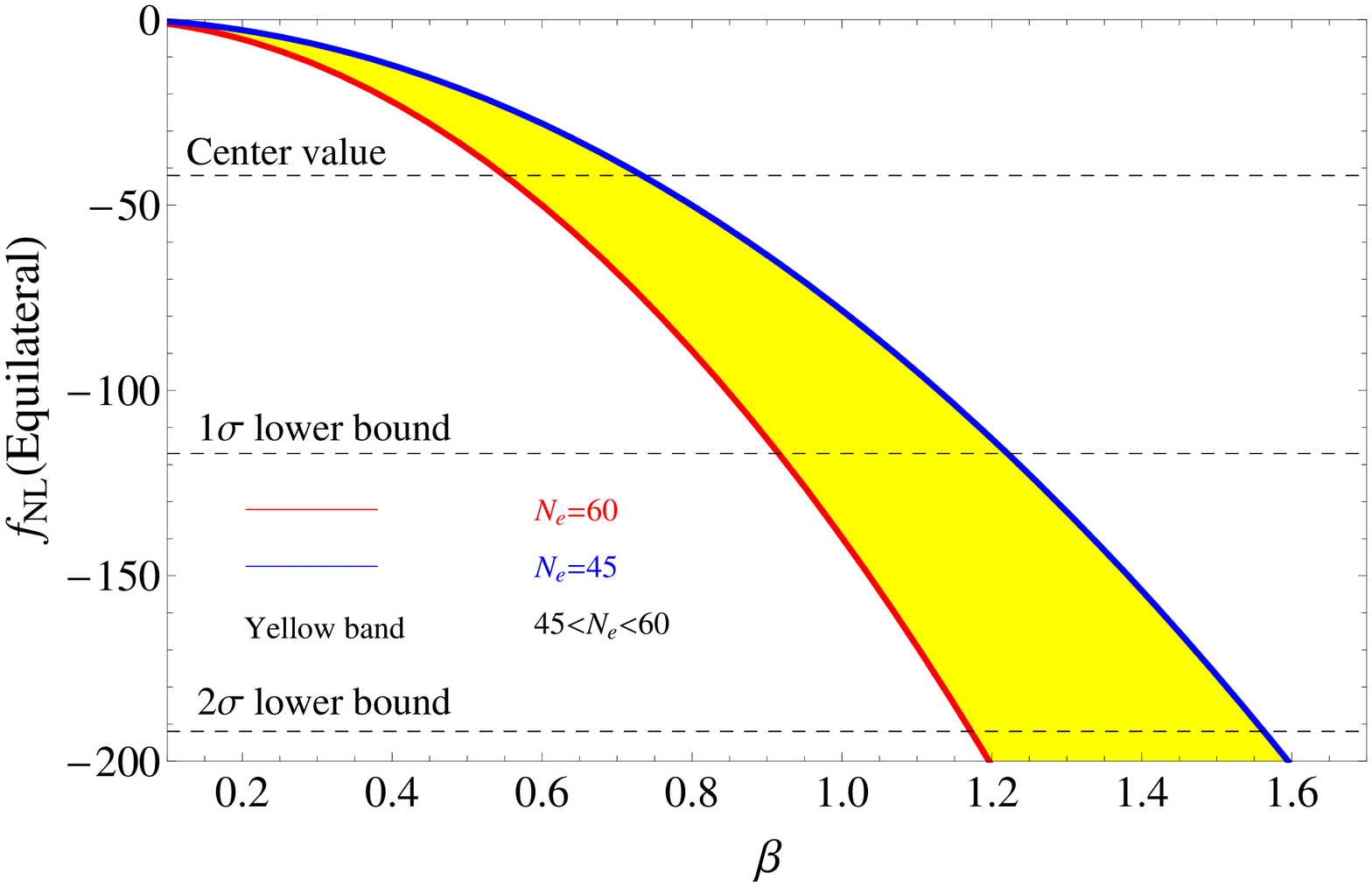}
\includegraphics[bb=0 0 624 410,
width=3.3in]{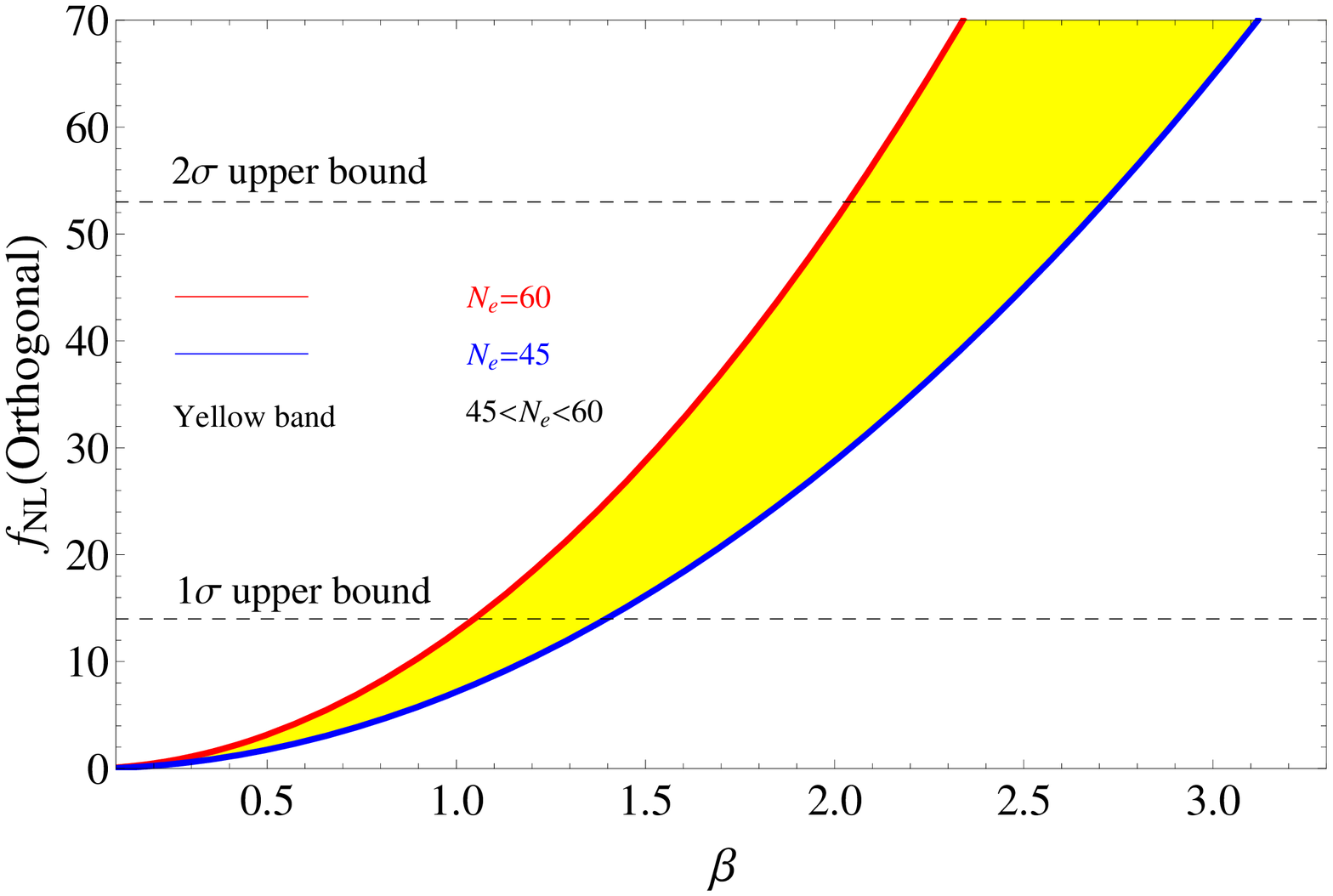}} \caption{Equilateral (\textit{left
panel}) and orthogonal (\textit{right panel}) $f_{\rm{NL}}$
predictions for different $\beta$ values in the IR DBI model. The
red and blue lines are for number of \textit{e}-folds to be $60$ and
$45$. The yellow band is the allowed region given these two boundary
of number of \textit{e}-folds. We over-plot the best-fit values,
$1\sigma$ and $2\sigma$ bounds for $f^{\rm{eq}}_{\rm{NL}}$ and
$f^{\rm{orth}}_{\rm{NL}}$ as dashed straight lines.}
\label{fig1:model3fnl}
\end{figure*}

\subsection{Model predictions}
In this section, we discuss another important type of brane
inflation model, namely, the infrared Dirac-Born-Infeld model (IR
DBI model). The difference between this model and the KKLMMT model
is that the rolling velocity of the brane is not determined by the
shape of the potential but by the speed limit of the warped
spacetime \cite{Bean08}. Such a warped spacetime can always emerge
in the inner space of compactified Calabi-Yau manifold.

Phenomenologically, the inflaton in IR DBI model can be driven by
the kinetic term, where the inflaton is not slowly rolling at all.
Therefore, the sound speed of inflaton in such a model could be
less than unity, providing a large tilt in the tensor power
spectrum (remember $n_{t}=-r/(8 c_{s})$ \cite{Ma09,Cheng12}).
Observation on large scale temperature and polarization can be
used to pin down the uncertainty of the soundspeed. In addition,
as shown in previous analyses \cite{Bean08}, there are a lot of
parameters that describe the structure of internal space, and we
will show that some of them may be pinned down by the CMB
observations (see also \cite{Planck13-22} for more detailed
discussions on constraints from non-Gaussianity).

In the DBI inflation, the action takes the form
\begin{equation}
P(\phi,X)=-f(\phi)^{-1}\sqrt{1-2f(\phi)X}+f(\phi)^{-1}-V(\phi),
\end{equation}
where $V(\phi)$ is the potential, $X$ is the kinetic term, and
$f(\phi)$ is the warp factor. For the IR DBI model, the inflaton
potential is
\begin{equation}
V(\phi)=V_0-{1\over 2}\beta H^2\phi^2,
\end{equation}
where the parameter $\beta$ is in principle within a wide range
$0.1<\beta<10^9$~\cite{Bean08}.

The scalar power spectrum of DBI inflation can be parametrized as
\cite{Bean08}
\begin{eqnarray}
\Delta^2_{\cal R}(k)=  \frac{A_{s}}{N^{4}_{e}}
\left(1-\frac{N^{16}_{c}}{N^{8}_{c}+(N^{\rm{DBI}}_{e})^{8}} \right),
\label{eq:DBIPk}
\end{eqnarray}
where $A_{s}$ is the amplitude of the perturbations which depends
on several parameters of the internal space, $N_{c}$ is the
critical number of \textit{e}-folds at scale $k_{c}$ (critical
scale where string phase transition happens), and
$N^{\rm{DBI}}_{e}$ is the number of \textit{e}-folds of inflation
at relativistic rolling. The total number of \textit{e}-folds is
the sum of relativistic and nonrelativistic (NR) rollings,
\begin{equation}
N^{\rm{tot}}_{e}=N^{\rm{DBI}}_{e}+N^{\rm{NR}}_{e}. \label{eq:Ntot}
\end{equation}
Now we can calculate the spectral index and its running, which
turns out to be (see also Appendix in \cite{Bean08})
\begin{eqnarray}
n_{s}-1 &=& \frac{d \ln \Delta^2_{\cal R}(k)}{d \ln k} \nonumber \\
&=& \frac{4}{N^{\rm{DBI}}_{e}} \frac{x^{2}+3x-2}{(x+1)(x+2)},
\nonumber \\
\alpha_{s} &=& \frac{d n_{s}}{d \ln k} \nonumber \\
&=&\frac{4}{(N^{\rm{DBI}}_{e})^{2}}
\frac{x^4+6x^3-55x^{2}-96x-4}{(x+1)^{2}(x+2)^{2}} ,
\end{eqnarray}
where $x=(N^{\rm{DBI}}_{e}/N_{c})^{8}$.

In addition, nontrivial sound speed $c_{s}$ can generate large
non-Gaussianity since the inflaton is no longer slowly rolling
down to the potential. The predicted equilateral and orthogonal
non-Gaussianities are \cite{Chen07JCAP,Senatore10}
\begin{eqnarray}
f^{\rm{eq}}_{\rm{NL}}=-0.35 \frac{1-c^{2}_{s}}{c^{2}_{s}},
\nonumber \\
f^{\rm{orth}}_{\rm{NL}}=0.032 \frac{1-c^{2}_{s}}{c^{2}_{s}},
\end{eqnarray}
where
\begin{eqnarray}
\frac{1}{c_{s}} \simeq \frac{\beta N^{\rm{DBI}}_{e}}{3}.
\label{eq:cs}
\end{eqnarray}

\subsection{Confront with current data}

Since the IR DBI model has a lot of parameters that describe the
internal structure of the warped space, in order to directly
compare its predictions with the current observational data, we
adopt the best-fit values of $N_{c}$, $k_{c}$ and
$N^{\rm{NR}}_{e}$ to be $35.7$, $10^{-4.15}~\textrm{Mpc}^{-1}$ and
$18.4$, respectively, according to the constraints from
\textit{WMAP} 5-year data \cite{Bean08}.

In Fig.~\ref{fig1:model3asns} we plot the predicted trajectory of
the IR DBI model in the $\alpha_{s}$--$n_{s}$ plane. The purple
and orange contours are the results from \textit{WMAP}9+ and
\textit{Planck}+WP+BAO as we discussed before. The red line is the
trajectory corresponding to $N^{\rm{tot}}_{e}$ between $45$ and
$60$, which includes a wide range of scale $k$. One can see that
the trajectory crosses the contours of both \textit{WMAP}9+ and
\textit{Planck}+WP+BAO, which is quite consistent with the data.
In addition, the model predictions at the two pivot scales
$k_{0}=0.002\textrm{Mpc}^{-1}$ and $k_{0}=0.038\textrm{Mpc}^{-1}$,
which are the chosen scales of the two constraints are marked on
the plot. One can see that the black dot is close to the boundary
of \textit{WMAP}9+ constraints while the blue one is outside of
the $2\sigma$ contours from \textit{Planck}. However, although it
seems that there is a discrepancy, we remind the reader that there
is some uncertainty of the subsequent evolution after inflation,
so it is reasonable to allow a broader range of number of
\textit{e}-folds for a given pivot scale.

Non-Gaussianity becomes an important tool to constrain such a
non-slow-roll inflation model. The local, equilateral and
orthogonal $f_{\rm{NL}}$ parameters are given by
Eq.~(\ref{nGnumber}), which still do not show strong signal for
non-Gaussianity. However, the error-bars of local, equilateral
(and orthogonal) $f_{\rm{NL}}$ become a factor of two and four
smaller than \textit{WMAP} 9-year data \cite{Bennett12}. Since the
IR DBI model predicts vanishing local $f_{\rm{NL}}$, it is already
consistent with the value given by \textit{Planck}. Now we
investigate the predictions of equilateral and orthogonal types of
non-Gaussianity.

In Fig.~\ref{fig1:model3fnl}, we plot the model predictions of
$f^{\rm{eq}}_{\rm{NL}}$ and $f^{\rm{orth}}_{\rm{NL}}$ and the
current lower and upper bounds. The yellow bands in both panels are
the allowed region for $N^{\rm{tot}}_{e}$ in between $45$ and $60$.
Note that in \textit{Planck} paper XXII \cite{Planck13-22}, $N_{e}$
is just allowed to be 60--90 when considering the constraints on the
IR DBI model, while here we consider a more reasonable range of the
number of \textit{e}-folds. Given the yellow bands and the $2\sigma$
lower bound for equilateral type of non-Gaussianity, we find that
the value of $\beta$ needs to be smaller than $1.5$ in order to
prevent very negative equilateral non-Gaussianity. Similarly, on the
right panel, we show that $\beta$ needs to be less than $2.5$ in
order to prevent large positive non-Gaussianity. These limits are
consistent with the range of $\beta<0.7$ as found by \textit{Planck}
paper XXIV \cite{Planck13-24}, which uses global likelihood analysis
to obtain the limit. We should notice that this is already a
fine-tuning for IR DBI model, because in this model $\beta$ has the
lower limit $0.1$ ($\beta<0.01$ is KKLMMT model as discussed in
Sec.~\ref{sec:kklmmt}) but no real upper limit. Therefore, the
current data is able to shrink the parameter space to be
$[0.1,\mathcal{O}(1)]$ is already a tight limit. Our comparison
gives a intuitive understanding of why the parameter $\beta$ needs
to be smaller than a certain value.

\section{Conclusion}
\label{sec:conclusion}

In this paper, we studied brane inflation with the \textit{Planck}
data and the joint data set from \textit{WMAP} 9-year data, SPT,
ACT, BAO and $H_{0}$ data. We first discussed the relationship
between the number of \textit{e}-folds and the corresponding pivot
scale. We clarified the case where adopting different pivot scales
of the constraints, the corresponding number of \textit{e}-folds
could be slightly different.

We then considered a toy model (prototype) of brane inflation
where the problem of dynamic stabilization is neglected.
Furthermore, we considered a more realistic ``slow-roll'' brane
inflation model (namely, the KKLMMT model) and the DBI inflation
model, and examined them with the \textit{Planck} and
\textit{WMAP}9+ results.

For the toy model, we showed that the model is consistent with the
observational data at $2\sigma$ CL, given the fact that it prefers
a red tilt close to $0.96$ and a slightly negative running. For a
comparison, in our previous work \cite{Ma09}, we found that this
type of brane inflation model is consistent with the \textit{WMAP}
5-year data at the level of $1\sigma$. The situation does not
change very much when we confront the model with \textit{WMAP}9+
data and \textit{Planck} data.

For the KKLMMT model, we first discussed how the model parameter
$\beta$ affects its predictions of scalar power spectrum. Then we
compared the model to the \textit{WMAP}9+ data and \textit{Planck}
data. We found that in order for the model to provide the
$\alpha_{s}$ and $n_{s}$ allowed by the tight constraints from
\textit{Planck} and \textit{WMAP}9+, the $\beta$ parameter needs
to be fine-tuned to be less than $10^{-3}$. For comparison, by
using the \textit{WMAP} 3-year data in \cite{Huang06}, we found
that the KKLMMT model cannot fit \textit{WMAP}3+SDSS data at the
level of 1$\sigma$ and a fine-tuning, at least eight parts in a
thousand, is needed at the level of $2\sigma$. When the
\textit{WMAP} 5-year data becomes available, we found that the
value of the parameter $\beta$ is constrained to be less than
${\cal O}(10^{-2})$ at the level of 2$\sigma$ \cite{Ma09}. Thus,
we can see that the problem of fine-tuning of $\beta$ becomes more
severe when confronting with the recent observational data.
Undoubtedly, this is not good news for the KKLMMT model.

Finally, we briefly discussed the current constraints on the
infrared Dirac-Born-Infeld inflation model given the current
observational data. The model can predict a larger negative
running ($\sim -0.02$) than the previous KKLMMT model. By figuring
out the trajectory of the model on the $\alpha_{s}$--$n_{s}$ plane
by varying the number of \textit{e}-folds, we found that the model
can predict the running of the spectral index and the tilt that
are consistent with \textit{WMAP}9+ and \textit{Planck} data.
However, when we confronted it with the current bounds on
equilateral and orthogonal non-Gaussianities, we found that in
order to avoid a large non-Gaussianity the value of $\beta$ which
controls the shape of the potential needs to be less than $1.5$.
This limit to the IR DBI model is already a fine-tuning.

To summarize, although the prototype of brane inflation can fit
the data well, it is not a realistic model of the brane inflation.
For the KKLMMT and IR DBI inflation models, the parameters need to
be fine-tuned to satisfy the current observational requirement.
The current observation of CMB from \textit{Planck} is competent
to place stringent limits on internal parameters of warped space.

\section*{Acknowledgments}
We would like to thank Anthony Challinor, Xingang Chen, Gary
Hinshaw and Andrew Liddle for useful discussions. Y.Z.M. is
supported by a CITA National Fellowship. Part of the research is
supported by the Natural Science and Engineering Research Council
of Canada. Q.G.H. is supported by the Knowledge Innovation Program
of the Chinese Academy of Science and by the National Natural
Science Foundation of China (Grant No.~10821504). X.Z. is
supported by the National Natural Science Foundation of China
(Grants No.~10705041, No. 10975032 and No. 11175042) and by the
National Ministry of Education of China (Grants No.~NCET-09-0276,
No. N100505001 and No. N120505003).

\end{document}